\documentstyle[prl,aps,amsmath,epsf,twocolumn]{revtex}

\begin{document}
\draft
\twocolumn[\hsize\textwidth\columnwidth\hsize\csname@twocolumnfalse%
\endcsname

\preprint{}

\title{Electrical and Thermal Transport by Nodal
Quasiparticles in the DDW State}
\author{Xiao Yang and Chetan Nayak}
\address{Physics Department, University of California, Los Angeles, CA 
  90095--1547}
\date{\today}
\maketitle

\begin{abstract}
We compute the electrical and thermal
conductivities and Hall conductivities of
the $d$-density wave (DDW) state in
the low-temperature impurity-scattering-dominated
regime for low-dopings, at which they are dominated by
nodal quasiparticles. We show that the
longitudinal conductivity in this limit in the DDW
state is not Drude-like. However, the thermal conductivty is
Drude-like; this is a reflection of the discrepancy between
electrical and thermal transport at finite frequency
in the DDW state.
An extreme example of this occurs in
the $\mu=0$, $\tau\rightarrow\infty$ limit, where
there is a strong violation of the Wiedemann-Franz
law: ${\kappa_{xx}}/{\sigma_{xx}} \propto {T^2}$
at $\omega=0$ and ${\kappa_{xx}}/{\sigma_{xx}}=0$ at
finite frequency. The DDW electrical and
thermal Hall conductivities are
linear in the magnetic field, $B$, for weak fields.
The formation of Landau levels at the nodes
leads to the quantization of these Hall
conductivities at high fields.
In all of these ways, the quasiparticles of the DDW
state differ from those
of the $d_{{x^2}-{y^2}}$ superconducting (DSC) state.
\end{abstract}
\vspace{1 cm}

\vskip -0.4 truein
]
\narrowtext

\section{Introduction.} 
Transport measurements have produced a wealth
of information about the high-$T_c$ superconducting
cuprates \cite{Ginsberg89}. However, it is not immediately clear how this
information can be used to shed light on their phase
diagram. When inelastic processes determine transport
coefficients, they are usually constrained by
the very phase space restrictions which underlie
Fermi liquid theory. Elastic scattering by impurities usually
leads to a temperature- and frequency-independent scattering
rate which is also not strongly dependent on the state of matter.
Thus, transport coefficients need not bear a very strong imprint
of the underlying phase. In more technical parlance, we would
say that transport coefficients are usually
detemined by irrelevant operators,
not by the fixed points themselves, except when they are
determined by impurity scattering which drives most systems
to a diffusive fixed point. However, the usual situation
does not always hold, and
in this paper we study one example where it does not.
We show that the properties of nodal quasiparticles
bespeak the broken symmetries which lead to their existence.
Consequently, low-temperature transport in the highly
underdoped limit (where it might be dominated by nodal
excitations) can be used to reveal
the physics of the pseudogap state of the underdoped cuprates.

\begin{figure}[thb]
\centerline{\epsfxsize=2.5in\epsffile{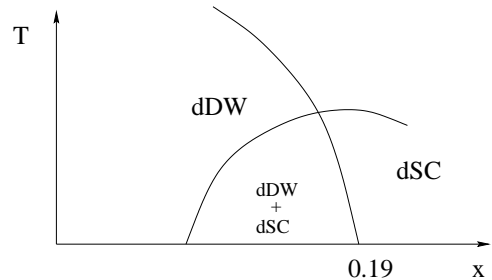}}
\vskip 0.5 cm
\caption{Schematic proposed phase diagram
for the cuprate superconductors.} 
\label{fig:phase}
\end{figure}

In \cite{Chakravarty01}, it was proposed that
the pseudogap phenomenon \cite{Timusk99} in high-$T_c$ cuprates
is the result of the development of another
order parameter called $d$-density wave order (DDW),
\begin{equation}
\label{eqn:order}
\left\langle {c^{\alpha\dagger}}({\bf k+Q},t)\,
{c_\beta}({\bf k},t) \right\rangle =
i{\Phi_{\bf Q}}\,f({\bf k})\,\,
{\delta^\alpha_\beta},
\end{equation}
where $f({\bf k})=\cos{k_x}a - \cos{k_y}a$.
This order parameter breaks the symmetries
of time-reversal, translation by
one lattice spacing, and rotation by $\pi/2$,
but respects the combination of any two
of these. There is no modulation of the charge density
in this state since $f({\bf k})$ vanishes upon
integration about the Fermi surface. However,
there are circulating currents in the ground state,
which alternate from one plaquette to the next
at wavevector ${\bf Q}=(\pi/a,\pi/a)$.
The underdoped superconducting state
was conjectured to exhibit both DDW
order and $d_{{x^2}-{y^2}}$ superconducting (DSC)
order. For hole dopings larger than a critical
value ${x_c}\approx 0.19$, DDW order is
presumed to disappear.
The proposed phase diagram is indicated
schematically in figure \ref{fig:phase}.
As a result of its $d_{{x^2}-{y^2}}$
angular variation, DDW order bears some similarities
to DSC order in its quasiparticle spectrum.
However, as we show in this paper, the transport
properties of these two phases are
rather different.

The development of DDW order forms the basis for
a compelling picture of the pseudogap because it
qualitatively explains many of the hallmarks of
pseudogap behavior such as the existence of a 
$d_{{x^2}-{y^2}}$ single-particle gap in photoemission
\cite{Norman98,Harris96}
and tunneling \cite{Renner98}, resonant inelastic neutron scattering
(``the 41 meV peak''), $c$-axis transport \cite{Homes93},
and the doping
dependence of the superfluid density \cite{Tallon99}.
The basic physical
picture underlying this description of the cuprates
involves the interplay between two order parameters
-- one superconducting and one density wave -- similar
to that observed in the A15 \cite{Bilbro76}
and dichalcogenide \cite{Vescoli98,Neto00} materials.

This consonance between theory and experiment
is encouraging but far from conclusive. However,
recent attempts to directly detect the DDW order parmeter have the
potential to be definitive. They have
centered on the observation of the small
magnetic fields ($\approx 10$G) which are generated
by the circulating currents which appertain to DDW
order. {\em Elastic} neutron scattering experiments
\cite{Mook01} on YBCO6.6 have found Bragg peaks in
spin-flip scattering at wavevector ${\bf Q}=(\pi/a,\pi/a)$,
with an onset temperature which is consistent with
that determined by other probes of the pseudogap.
These experiments thereby observe the symmetry-breaking pattern
associated with DDW order. Antiferromagnetism also
breaks time-reversal and translation by one lattice
spacing, but the form factors measured in the experiment
are more consistent with DDW order.
A recent $\mu$SR experiment \cite{Sonier} finds evidence for
small time-reversal symmetry breaking in YBCO.
An exciting feature of these experiments is that
the onset temperature is broadly consistent with
the pseudogap scale in YBCO6.67 and the onset temperature
is {\em below} $T_c$ in YBCO6.95, as would be expected from
a phase diagram of the type depicted in figure \ref{fig:phase}.

Can in-plane transport shed further light on the
nature of the pseudogap state? In this paper, we show
how nodal quasiparticles evince the symmetries of the phase
which supports them by contrasting the transport properties of
quasiparticles in the $d_{{x^2}-{y^2}}$-density wave (DDW) and
$d_{{x^2}-{y^2}}$ superconducting (DSC) phases. We show how
the former reflect the broken translational symmetry of the DDW phase.
We find a dichotomy between electrical and thermal transport;
nodal quasiparticles in the DDW state carry
electrical current much more effectively than
they carry thermal current. A further distinction
stems from the fact that the DDW state, unlike the DSC
state, does not break gauge invariance, so the quasiparticles
of the DDW state form Landau levels and
have Hall electrical or thermal conductivities which are linear in
the magnetic field, $B$, in contrast to their DSC cousins.
These distinctions are sharpest in the limit of small
$\mu$, where the quasiparticle excitations of the
DDW state are restricted to the vicinity of the nodal
points. In this limit, the quasiparticles of the DDW state
are analogous to `relativistic' electrons and positrons.
The quasiparticles of the DSC state, on the other hand,
couple differently to the electromagnetic field
as a result of their coherence factors
which mix particle and hole states.

In section II, we briefly describe the
low-energy effective field theories for quasiparticles
in the DDW state, the DSC state, and the state
with both DDW and DSC order. In
section III, we use gauge invariance and Noether's
theorem to derive the electrical and energy current operators
in these low-energy effective field theories.
In sections IV and V, we present and discuss
our results for these conductivities in the three different
phases considered. Finally, in section VI, we discuss
our results and consider their possible range of validity.
In appendix A, we discuss the Kubo formulae which
relate the electrical and thermal conductivities and
Hall conductivities to correlation functions of the
appropriate current operators. In appendix B,
we review impurity scattering
of nodal quasiparticles.

\section{Nodal Quasiparticle Hamiltonian}

The low-energy quasiparticle Hamiltonian for
the DDW state is:
\begin{eqnarray}
{H^{\rm DDW}} &=& \int\frac{d^{2}k}{(2\pi)^{2}}
[\left(\epsilon(k)-\mu\right)c^{\alpha\dagger}(k)c_{\alpha}(k) +\cr
& & {\hskip 2 cm}
i{\Delta(k)}c^{\alpha\dagger}(k)c_{\alpha}(k+Q)]
\end{eqnarray}
where $\epsilon(k)$ is the single-particle energy
which we can, for instance, take to be
\begin{eqnarray}
\label{eqn:band-structure}
\epsilon(k) = -2t(\cos{k_x}a + \cos{k_y}a)
+ 4t' \cos{k_x}a \cos{k_y}a
\end{eqnarray}
while
\begin{eqnarray}
\label{eqn:gap-structure}
\Delta(k) = \frac{\Delta_{0}}{2}(\cos k_{x}a - \cos k_{y}a)
\end{eqnarray}
is the d-wave order parameter of the DDW state.
${\bf Q} = (\frac{\pi}{a}, \frac{\pi}{a})$.

Since the order parameter breaks translational symmetry
by one lattice spacing, it is convenient to
halve the Brillouin zone and form a two-component
electron operator:
\begin{eqnarray}
\label{eqn:trans-indices}
\left( \begin{array}{c}
            \chi_{1\alpha} \\ \chi_{2\alpha}
           \end{array} \right) = \left( \begin{array}{c}
             c_{\alpha}(k) \\ i c_{\alpha}(k+Q)
                                         \end{array} \right)
\end{eqnarray}
Then the mean field Hamiltonian in terms of $\chi$ becomes 
\begin{eqnarray}
H &=& \int\frac{d^{2}k}{(2\pi)^{2}}\,\chi^{\alpha\dagger}(k)
\biggl[\frac{1}{2}(\epsilon(k) + \epsilon(k+Q)) - \mu\cr
& & \frac{1}{2}(\epsilon(k) - \epsilon(k+Q))\sigma^{(3)} \:+\:
\Delta(k)\sigma^{(1)}\biggr]\chi_{\alpha}(k)
\end{eqnarray}
Where $\vec{\sigma}$ are Pauli matrices which mix
the two translational components of (\ref{eqn:trans-indices}).

The single-quasiparticle energy is:
\begin{eqnarray}
E_{\pm}(k) &=& \frac{1}{2}(\epsilon(k) + \epsilon(k+Q))\cr
& & \pm\: \frac{1}{2}\sqrt{(\epsilon(k) - \epsilon(k+Q))^{2} + 4\Delta^{2}(k)}
\end{eqnarray}
At exactly half-filling, there are 4 nodal points
$(\pm\frac{\pi}{2a}, \pm\frac{\pi}{2a})$
at which there are gapless excitations.
In the vicinity of these `Dirac points', the
spectrum is conical.
When the chemical potential is less than 0,
the states will be filled up to ${E_-}(k)=\mu$,
so the nodes will open into small pockets
which are cross sections of the `Dirac cone'.
The low-energy physics will be dominated by these 
gapless fermionic excitations.
We can focus on a single pair of nodal points,
$(\frac{\pi}{2a},\frac{\pi}{2a})$ and
$(-\frac{\pi}{2a},-\frac{\pi}{2a})$.
It is straightforward to include the other pair
of nodes into our result at the end of any calculation.
We choose the x-axis to be
perpendicular to the free-electron Fermi surface and the
y-axis to be parallel to the free-electron Fermi
surface at one antipodal pair of nodes; the x-axis is
parallel to the free-electron Fermi surface
and the y-axis is perpendicular to the free-electron
Fermi surface at the other pair.

\begin{figure}[thb]
\centerline{\epsfxsize=2.3in\epsffile{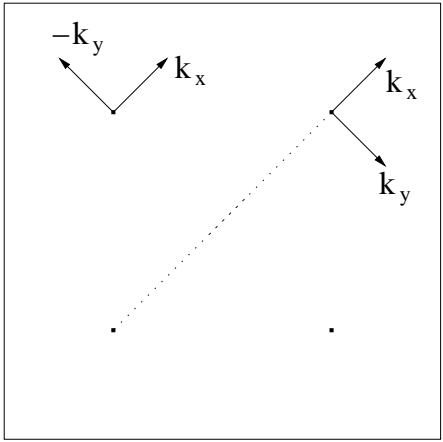}}
\vskip 0.5 cm
\caption{The quasiparticle spectra are linearized about
the Fermi points $(\pm\pi/2,\pm\pi/2)$. We choose the
$x$ and $y$ axes as shown. Each pair of antipodal nodes
(shown connected by dotted lines) is combined to
form a single Dirac fermion.} 
\label{fig:fermi-points}
\end{figure}

Linearizing the spectrum about the nodes,
we obtain the linearized dispersion relation:
\begin{equation}
E(k) = \pm\sqrt{v^{2}_{F}k^{2}_{x} + 
v^{2}_{\Delta^{DDW}}k^{2}_{y}}
\end{equation}
where ${v_F}=2\sqrt{2}\,ta$ and
$v^{\rm DDW}_{\scriptscriptstyle\Delta} = \sqrt{2}\Delta_{0}a$;
$t'$ does not enter at linear order.
The effective Lagrangian of the nodal quasiparticles is:
\begin{eqnarray}
\label{eqn:DDW-linear-action}
{\cal L}^{DDW}_{eff} &=& \chi_\alpha^{1\dagger}(\partial_{\tau} + \mu -
v_{F}\sigma^{(3)}i\partial_{x} -
v^{\rm DDW}_{\scriptscriptstyle\Delta}
\sigma^{(1)}i\partial_{y})\chi^1_{\alpha}\cr
&+& \chi_\alpha^{2\dagger}(\partial_{\tau} + \mu -
v^{\rm DDW}_{\scriptscriptstyle\Delta}\sigma^{(3)}i\partial_{x} -
v_{F}\sigma^{(1)}i\partial_{y})\chi^2_{\alpha}
\end{eqnarray}
The superscript $1,2$ labels the two antipodal pairs of
nodal points.
Note that the chemical potential, $\mu$, moves
the free-electron Fermi surface away from the nodes and opens
Fermi pockets which are the sections of the Dirac cone
referred to earlier. When $\mu$ is small, the Fermi surface will
remain in the vicinity of the nodes and we
can use the linearized effective action (\ref{eqn:DDW-linear-action}).
In this paper, we will work in the linearized approximation;
in the discussion, we will examine the regime of
validity of this approximation.
The associated nodal quasiparticle Green functions are
\begin{eqnarray}
{G^1}({i\epsilon_n},{\bf k}) = \frac{i{\epsilon_n}+\mu
+ {\sigma_z}{v_{F}}{k_x}+{\sigma_x}
{v^{\rm DDW}_{\scriptscriptstyle\Delta}}{k_y}}{{({i\epsilon_n}+\mu)^2}
-{v_{F}^2}{k_x^2}-
{\left({v^{\rm DDW}_{\scriptscriptstyle\Delta}}\right)^2}{k_y^2}}
\end{eqnarray}
and a similar expression for ${G^2}({i\epsilon_n},{\bf k})$
with ${v_{F}}$ and ${v^{\rm DDW}_{\scriptscriptstyle\Delta}}$ exchanged.

Now consider the DSC state. We can write the
low-energy quasiparticle Hamiltonian as:
\begin{eqnarray}
{H^{\rm DSC}} &=& \int \frac{{d^2}k}{(2\pi)^2}\:
{\psi_\alpha^\dagger}(k) \biggl[
{\tau^{(3)}}\left(\epsilon\left({\bf k}\right)-\mu\right)
+ {\tau^{(1)}} \Delta(k)\biggr]{\psi_\alpha}(k)
\end{eqnarray}
where
\begin{equation}
\left[ \begin{array}{c}
\psi_{1\alpha}(\vec{k}) \\  
\psi_{2\alpha}(\vec{k})
\end{array}\right]
  = \left[ \begin{array}{c}
c_{k\alpha}^{\vphantom\dagger} \\  
{\epsilon_{\alpha\beta}}\,c_{-k\beta}^{\dagger} 
\end{array}\right].
\end{equation}
and the Pauli matrices $\tau$ act on the particle-hole indices
of this Nambu-Gorkov spinor.

We now linearize the spectrum about $(\pm\frac{\pi}{2a}, \pm\frac{\pi}{2a})$,
which is not the same as linearizing about
the free-electron Fermi surface $\epsilon(k)=\mu$ (which is more
conventional)
if $\mu \neq 0$. We obtain the effective action
for DSC quasiparticles:
\begin{eqnarray}
{\cal L}^{DSC}_{eff} &=& \psi_\alpha^{1\dagger}(\partial_{\tau} -
v_{F}\tau^{(3)}\left(i\partial_{x}-\mu\right) -
v^{\rm DSC}_{\scriptscriptstyle\Delta}\tau^{(1)}i\partial_{y}){\psi^1_\alpha}
\cr
& & + \:\:\left(1\rightarrow 2\:,\:x\leftrightarrow y\right)
\end{eqnarray}
Again, the superscript $1,2$ labels the pairs of nodes.
The Green functions are formally similar, although
the matrix labels have different meanings;
they label magnetic zones in the DDW
case and they label particles and holes in the
DSC case. An important difference between the
two states is that in the DSC case,
the chemical potential moves the nodes along
with the Fermi surface so that they are at ${k_x}=\mu/{v_F}$.

The difference in the way in which the chemical potential,
$\mu$, enters the DDW and DSC nodal quasiparticle Hamiltonians
is a harbinger of the difference in their electrical
responses. In the next section, we show how this
difference manifests itself in the form of the current
operators in the two states.

First, however, let us consider a state with both order parameters.
It is useful to define a four-component object, $\Psi$:
 \[ \left( \begin{array}{c}
            \Psi_{1} \\ \Psi_{2} \\ \Psi_{3} \\ \Psi_{4}
           \end{array} \right) = \left( \begin{array}{c}
          c_{\alpha}(k) \\ c_{\alpha}(k+Q) \\
{\epsilon_{\alpha\beta}} c^{\dagger}_{\beta}(-k) \\
{\epsilon_{\alpha\beta}} c^{\dagger}_{\beta}(-k-Q)
                                         \end{array} \right)  \]
Linearizing about the nodal points at
half-filling, $(\pm\frac{\pi}{2a}, \pm\frac{\pi}{2a})$,
we can write the effective action as:
\begin{eqnarray}
{\cal L}^{DDW+DSC}_{eff} &=& \Psi^{\dagger}\biggl(\partial_{\tau}+
\mu{\Gamma^c} -
v_{F}\Gamma^{1}i\partial_{x} \cr
& & {\hskip 0.5 cm}
- {v^{\rm DSC}_{\scriptscriptstyle\Delta}}\Gamma^{2}i\partial_{y}
- {v^{\rm DDW}_{\scriptscriptstyle\Delta}}\Gamma^{3}i\partial_{y}\biggr)\Psi
\end{eqnarray}
where
\[ \Gamma^{c}=\left( \begin{array}{lccr}
                      I & 0 \\
                      0 & -I \\
                      \end{array} \right) \]
\[ \Gamma^{1}=\left( \begin{array}{lccr}
                      {\sigma^3} & 0 \\
                      0 & -{\sigma^3} \\
                      \end{array} \right) \]
\[ \Gamma^{2}=\left( \begin{array}{lccr}
                      0 & {\sigma^1} \\
                      {\sigma^1} & 0 \\
                      \end{array} \right) \]
\[ \Gamma^{3}=\left( \begin{array}{cc}
                      {\sigma^2} & 0 \\
                      0 & -{\sigma^2} \\
                     \end{array} \right) \]

\section{Electrical and Energy Current Operators}

Although the DSC and DDW states have similar quasiparticle
spectra near their nodal points, the coupling of these
quasiparticles to the electromagnetic field is different.
As a result, the electrical current operators are
different in the two states.

In the DSC state, the electrical current is:
\begin{eqnarray}
\label{eqn:dsc-current-full}
{j_x} &=& -e{v_F}\,{\psi^{1\,\dagger}_{\alpha}}{\psi^{1}_{\alpha}}\cr
{j_y} &=& e{v_F}\,{\psi^{2\,\dagger}_{\alpha}}{\psi^{2}_{\alpha}}
\end{eqnarray}
At a given node, there is no current parallel to the Fermi
surface; hence, there is no contribution of
${\psi^{1}_{\alpha}}$ to $j_y$ and no contribution
of ${\psi^{2}_{\alpha}}$ to $j_x$. In essence,
the DSC order parameter does not affect the quasiparticle
current.

In the DDW state, the elctrical current is given by
\begin{eqnarray}
\label{eqn:ddw-current-full}
{j_x} &=& -e{v_F}\,{\chi^{1\,\dagger}_{\alpha}}
{\sigma^{(3)}}{\chi^{1}_{\alpha}} \:+\:
e{v_{\scriptscriptstyle\Delta}}\,{\chi^{2\,\dagger}r_{\alpha}}
{\sigma^{(1)}}{\chi^{2}_{\alpha}}\cr
{j_y} &=& -e{v_{\scriptscriptstyle\Delta}}\,
{\chi^{1\,\dagger}_{\alpha}}{\sigma^{(1)}}{\chi^{1}_{\alpha}}
\:+\:
e{v_F}\,{\chi^{1\,\dagger}_{\alpha}}{\sigma^{(3)}}{\chi^{1}_{\alpha}}
\end{eqnarray}
Both nodes contribute to the current operators in both directions.
As a result of the DDW order parameter, the quasiparticle
current operator is modified from its free-particle
form.

However, the energy current operator is identical in
form in the two states:
\begin{eqnarray}
\label{eqn:both-energy}
{j^E_x} &=& \frac{i}{2}v_{F}(\chi^{1\,\dag}\sigma^{(3)}\dot{\chi}^{1} - h.c)
- \frac{i}{2}v_{\scriptscriptstyle\Delta}
(\chi^{2\,\dag}\sigma^{(1)}\dot{\chi}^{2} - h.c)\cr
{j^E_y} &=& \frac{i}{2}v_{\scriptscriptstyle\Delta}
(\chi^{1\,\dag}\sigma^{(1)}\dot{\chi}^{1} - h.c)
- \frac{i}{2}v_{F}(\chi^{2\,\dag}\sigma^{(3)}\dot{\chi}^{2} - h.c)
\end{eqnarray}
where $v_{\scriptscriptstyle\Delta}$ is either 
$v^{\rm DDW}_{\scriptscriptstyle\Delta}$ or
$v^{\rm DSC}_{\scriptscriptstyle\Delta}$ and $\chi$
is replaced by $\psi$ in the latter case.

To see how (\ref{eqn:dsc-current-full})-(\ref{eqn:both-energy})
arise, first consider
the quasiparticle ``kinetic energy'' in the DSC state:
\begin{eqnarray}
H_{DSC} &=& \int {d^2}x\: \psi^{\dagger}
v_{F}\tau^{(3)}\left(i\partial_{x}-\mu\right)\psi\:\:+\cr 
& & {\hskip -1 cm} \int {d^2}x {d^2}x'\: {\Delta^{\rm DSC}}(R,x-x')\:
\psi^{\dagger}(x)\tau^{(1)}\psi(x')
\end{eqnarray}
In the second line, we have written the gap,
${\Delta^{\rm DSC}}$, which is a function of the
two electron coordinates, $x$ and $x'$, in terms of the
center-of-mass coordinate, $R$, and the relative coordinate,
$x-x'$. The $\tau^{(i)}$ are Pauli matrices.

We should insert the electromagnetic field into
this Hamiltonian in order to make it invariant
under the gauge transformation
\begin{eqnarray}
\label{eqn:gauge-trans1}
\psi &\rightarrow&
{e^{i \tau^{(3)}\varphi(x)}}\,\psi\cr
{\Delta^{\rm DSC}}(R,x-x') &\rightarrow&
{e^{-2i \varphi(R)}} \,{\Delta^{\rm DSC}}(R,x-x') 
\end{eqnarray}
The second line of (\ref{eqn:gauge-trans1})
is accurate to lowest order in $x-x'$, which
is sufficient since we will eventually
be linearizing about the nodes.
The first line of (\ref{eqn:gauge-trans1})
includes both of the transformation laws
\begin{eqnarray}
{c_\alpha}(x) &\rightarrow&
{e^{i \varphi(x)}}\,{c_\alpha}(x)\cr
{c^\dagger_\alpha}(x) &\rightarrow&
{e^{-i \varphi(x)}}\,{c^\dagger_\alpha}(x)
\end{eqnarray}

The dependence of the gap
on $x-x'$ determines the structure of
the pair wavefunction. In a $d$-wave superconductor
the two electrons in a pair cannot be on the same site.
For instance, if the gap varies as $\cos{k_x}a-\cos{k_y}a$,
the distance between the electrons, or the pair size,
is one lattice constant. When we linearize the gap
about the nodes, we treat this pair size as if it were
infinitesimal. In order to see how this works,
let us, as in the previous section, align the axes
so that the $x$-axis is perpendicular to the
Fermi surface and the $y$-axis is parallel to
the Fermi surface. Let us assume that the pairs have a fixed
size $a$ and that they are oriented along the
$\hat{\bf y}$ axis. Then, we linearize the gap
by making the following approximation:
\begin{eqnarray}
{\Delta^{\rm DSC}}({\bf R},{\bf x}-{\bf x}')\:
= {\hskip 4.5 cm}\cr
{v_{\scriptscriptstyle\Delta}}({\bf R})
\left(\frac{i}{2a}\right)\left[ \delta({\bf x}-{\bf x'}+a\hat{\bf y})
- \delta({\bf x}-{\bf x'}-a\hat{\bf y})\right]\cr
\approx {v_{\scriptscriptstyle\Delta}}({\bf R})
{\Delta^{\rm DSC}}({\bf R})\,i{\partial_y}
\end{eqnarray}
Under a gauge transformation, the gap transforms
as
\begin{eqnarray}
{v_{\scriptscriptstyle\Delta}}({\bf R})
\left(\frac{i}{2a}\right)\left[ \delta({\bf x}-{\bf x'}+a\hat{\bf y})
- \delta({\bf x}-{\bf x'}-a\hat{\bf y})\right] {\hskip 0.5 cm}\cr
\rightarrow \:
{e^{-2i \varphi({\bf R})}} \,
{v_{\scriptscriptstyle\Delta}}({\bf R})
\left(\frac{i}{2a}\right)\biggl[
\delta({\bf x}-{\bf x'}+a\hat{\bf y})
- {\hskip 0.5 cm}\cr
\delta({\bf x}-{\bf x'}-a\hat{\bf y})\biggr]\cr
 = {v_{\scriptscriptstyle\Delta}}({\bf R})
\left(\frac{i}{2a}\right)\biggl[
{e^{-2i \varphi({\bf x}+\frac{a}{2}\hat{\bf y})}}
\delta({\bf x}-{\bf x'}+a\hat{\bf y})\: -
{\hskip 0.2 cm}\cr
{e^{-2i \varphi({\bf x} - \frac{a}{2}\hat{\bf y})}}
\delta({\bf x}-{\bf x'}-a\hat{\bf y})\biggr]\cr
\approx {v_{\scriptscriptstyle\Delta}}({\bf R})
\delta({\bf x}-{\bf x'})\:
\frac{1}{2}\,i{\partial_y}{e^{-2i \varphi({\bf x})}}
\cr
\approx {v_{\scriptscriptstyle\Delta}}({\bf R})
\delta({\bf x}-{\bf x'})\:
{e^{-i \varphi({\bf x})}} 
i{\partial_y} {e^{-i \varphi({\bf x})}} 
{\hskip 1.9 cm}
\end{eqnarray}

In other words, if we linearize the Hamiltonian
about the nodes, as in the previous section,
\begin{eqnarray}
H_{DSC} &=& \int {d^2}x\: \psi^{\dagger}
v_{F}\tau^{(3)}\left(i\partial_{x}-\mu\right)\psi\cr 
& & + \: \int {d^2}x \: 
\psi^{\dagger}v^{\rm DSC}_{\scriptscriptstyle\Delta}\tau^{(1)}i\partial_{y}\psi
\end{eqnarray}
then the gap transforms as
\begin{eqnarray}
{v_{\scriptscriptstyle\Delta}} 
i{\partial_y} \rightarrow
{v_{\scriptscriptstyle\Delta}} {e^{-i \varphi({\bf x})}} 
i{\partial_y} {e^{-i \varphi({\bf x})}} 
\end{eqnarray}

Hence, the second (gap) term in the Hamiltonian
is already gauge-invariant, even without inserting
the electromagnetic field. In order to make
the Hamiltonian gauge-invariant,
we must couple the electromagnetic field in
the following way:
\begin{eqnarray}
H_{DSC} &=& \int {d^2}x\: \psi^{\dagger}
v_{F}\tau^{(3)}\left(i\partial_{x}+e{A_x}\tau^{(3)}-\mu\right)\psi\cr 
& & + \: \int {d^2}x \: \psi^{\dagger}
v^{\rm DSC}_{\scriptscriptstyle\Delta}\tau^{(1)}i\partial_{y}\psi
\end{eqnarray}

Consequently,
\begin{eqnarray}
{j_x} &=& -e{v_F}{\psi^\dagger_{\alpha A}}{\psi^{}_{\alpha A}}\cr
{j_y} &=& 0
\end{eqnarray}
in the DSC state. Combining this with the contribution
from the other set of nodes, we obtain
(\ref{eqn:dsc-current-full}).

Now consider the DDW effective Hamiltonian:
\begin{eqnarray}
{H_{\rm DDW}} &=& \int {d^2}x \, {\chi^\dagger_{\alpha A}}
{\sigma^{(3)}_{AB}} {v_F} i {\partial_x} {\chi_{\alpha A}}\:\: +\cr
& & {\hskip - 1 cm}\int {d^2}x\,{d^2}x'\,{\Delta^{\rm DDW}}(R,x-x')\,
{\chi^\dagger_{\alpha A}}\,{\sigma^{(1)}_{AB}}
 \,{\chi_{\alpha B}}
\end{eqnarray}
The $\sigma^{(i)}$ are Pauli matrices. We want this
effective Hamiltonian to be invariant under
\begin{eqnarray}
\label{eqn:gauge-trans2}
{\chi_{\alpha A}} &\rightarrow& {e^{i\varphi(x)}}\,{\chi_{\alpha A}}\cr
{\chi^\dagger_{\alpha A}} &\rightarrow& {e^{-i\varphi(x)}}\,
{\chi^\dagger_{\alpha A}}\cr
{\Delta^{\rm DDW}}(R,x-x') &\rightarrow& {\Delta^{\rm DDW}}(R,x-x')
\end{eqnarray}
The third line of (\ref{eqn:gauge-trans2})
is accurate to lowest order in $x-x'$, which
is sufficient since we will linearize about the nodes:
\begin{eqnarray}
{\Delta^{\rm DDW}}(R,x-x') \approx {v_{\scriptscriptstyle\Delta}}
\delta({\bf x}-{\bf x'})\,i{\partial_y}
\end{eqnarray}
Both terms in the linearized Hamiltonian
\begin{equation}
{H_{\rm DDW}} = \int {d^2}x \, {\chi^\dagger_{\alpha A}}
{\sigma^{(3)}_{AB}} {v_F} i {\partial_x} {\chi_{\alpha A}}
\:+\: {\chi^\dagger_{\alpha A}}
{\sigma^{(1)}_{AB}} {v_{\scriptscriptstyle\Delta}} i {\partial_y}
{\chi_{\alpha A}}
\end{equation}
transform under a gauge transformation, so the
electromagnetic field must be inserted in
both terms in the linearized Hamiltonian
\begin{eqnarray}
{H_{\rm DDW}} &=& \int {d^2}x \, {\chi^\dagger_{\alpha A}}
{\sigma^{(3)}_{AB}} {v_F} \left(i {\partial_x}+e{A_x}\right) {\chi_{\alpha A}}
\:+\cr
 & & \int {d^2}x \,{\chi^\dagger_{\alpha A}}
{\sigma^{(1)}_{AB}} {v_{\scriptscriptstyle\Delta}}
\left(i {\partial_y}+e{A_y}\right) {\chi_{\alpha A}}
\end{eqnarray}
Hence, the current operator in the DDW state
takes the form
\begin{eqnarray}
{j_x} &=& -e{v_F}{\chi^\dagger_{\alpha A}}
{\sigma^{(3)}_{AB}}{\chi^{}_{\alpha B}}\cr
{j_y} &=& -e{v_{\scriptscriptstyle\Delta}}{\chi^\dagger_{\alpha A}}
{\sigma^{(1)}_{AB}}{\chi^{}_{\alpha B}}
\end{eqnarray}
Combining this with the contribution from the
other node, we obtain (\ref{eqn:ddw-current-full}).

Now consider the state with both DDW and DSC order.
Its quasiparticles have effective action:
\begin{eqnarray}
{\cal L}^{\rm DDW+DSC}_{eff} &=& \Psi^{\dagger}\biggl(\partial_{\tau}+
\mu{\Gamma^\mu} -
v_{F}\Gamma^{1}i\partial_{x} \cr
& & {\hskip 0.5 cm}
- {v^{\rm DSC}_{\scriptscriptstyle\Delta}}\Gamma^{2}i\partial_{y}
- {v^{\rm DDW}_{\scriptscriptstyle\Delta}}\Gamma^{3}i\partial_{y}\biggr)\Psi
\end{eqnarray}
From the preceeding discussion of the two order parameters separately,
it is clear that the correct minimally-coupled form
of the action is:
\begin{eqnarray}
{\cal L}^{\rm DDW+DSC}_{eff} &=& \Psi^{\dagger}\biggl(\partial_{\tau}+
\mu{\Gamma^c} -
v_{F}\Gamma^{1}\left(i\partial_{x}+e{A_x}{\Gamma^c}\right) \cr
& & {\hskip - 1 cm}
- {v^{\rm DSC}_{\scriptscriptstyle\Delta}}\Gamma^{2}i\partial_{y}
- {v^{\rm DDW}_{\scriptscriptstyle\Delta}}\Gamma^{3}
\left(i\partial_{y}+e{A_y}{\Gamma^c}\right)\biggr)\Psi
\end{eqnarray}
from which it follows that the electrical current
is given by:
\begin{eqnarray}
j_{x}&=&-ev_{F}\Psi^{\dagger}{\Gamma^1}{\Gamma^c}\Psi\cr
j_{y}&=&-ev^{\rm DDW}_{\scriptscriptstyle\Delta}\Psi^{\dagger}
{\Gamma^3}{\Gamma^c}\Psi
\end{eqnarray}

In order to compute the thermal conductivity,
we will need the heat current.
The heat current ${j^Q}$ is related to the energy
current ${j^E}$ and the electrical current $j$ according to:
\begin{equation}
{{\bf j}^Q} = {{\bf j}^E} - \frac{\mu}{e}{\bf j}
\end{equation}
The thermal conductivity is measured under
a condition of vanishing electrical current flow,
${\bf j}=0$, so that ${{\bf j}^Q} = {{\bf j}^E}$.
The energy current may be obtained from Noether's
theorem. It is given by off-diagonal components of
the energy-momentum tensor:
\begin{equation}
{j^E_i} = {T_{0i}} = {\pi_i}{\partial_0}\phi
+ {\pi_i^\dagger}{\partial_0}{\phi^\dagger}
\end{equation}
where $\phi$ and $\phi^\dagger$ are the basic
fields in the theory and
${\pi_i}=\partial{\cal L}/\partial({\partial_i}\phi)$
are their canonical conjugates. (Note that $T_{i0}$ is
the momentum density, which can be distinct from
${T_{0i}}$, the energy current.)

In the DDW phase, this is:
\begin{eqnarray}
\label{eqn:ddw-energy}
{j^E_x} &=& \frac{i}{2}v_{F}(\chi^{\dag}\sigma^{(3)}\dot{\chi} - h.c)\cr
{j^E_y} &=& \frac{i}{2}v^{\rm DDW}_{\scriptscriptstyle\Delta}
(\chi^{\dag}\sigma^{(1)}\dot{\chi} - h.c)
\end{eqnarray}
while, in the DSC phase, it is identical in form:
\begin{eqnarray}
\label{eqn:dsc-energy}
{j^E_x} &=& \frac{i}{2}v_{F}(\psi^{\dag}\tau^{(3)}\dot{\psi} - h.c)\cr
{j^E_y} &=& \frac{i}{2}v^{\rm DSC}_{\scriptscriptstyle\Delta}
(\psi^{\dag}\tau^{(1)}\dot{\psi} - h.c)
\end{eqnarray}
When both order parameters are present, the energy current
is given by
\begin{eqnarray}
\label{eqn:ddw-dsc-energy}
{j^E_x} &=& \frac{i}{2}v_{F}(\Psi^{\dag}\Gamma^{1}\dot{\Psi} - h.c)\cr
{j^E_y} &=& \frac{i}{2}\left(
\psi^{\dag}\left(v^{\rm DSC}_{\scriptscriptstyle\Delta}\Gamma^{2} +
v^{\rm DDW}_{\scriptscriptstyle\Delta}\Gamma^{3}\right)\dot{\psi} - h.c\right)
\end{eqnarray}

We can use these current operators and the Green
functions derived in the previous section or in 
appendix B to compute electrical and thermal
copnductivities according to the Kubo formulae
given in appendix A.
In the next section, we will do this in the clean limit
in which there are no impurities. In section V, we will
consider the more realistic case in which the nodal
quasiparticles have a finite lifetime as a result of
impurity scattering.

\section{Electrical and Thermal Conductivities in the Clean Limit}

\subsection{DDW State}

It is instructive to consider the electrical and thermal
conductivities of the DDW state in the simplest
case imaginable: a completely pure
system with no impurities and no phonons. Let
us further neglect all interactions between the
nodal quasiparticles. Eventually, we will restore
impurity scattering and consider the low-temperature
limit in which impurity scattering dominates. However,
it is instructive to consider the ideal case first.

The frequency- and temperature-dependent electrical
conductivity consists of {\em two} contributions:
\begin{equation}
\sigma_{xx}(\omega,T) = \sigma_{xx}^{\rm Drude}(\omega,T) +
\sigma_{xx}^{\rm Inter}(\omega,T)
\end{equation}
The appellations `Drude' and `Inter' will
be explained shortly. These contributions
have the form
\begin{eqnarray}
\sigma_{xx}^{\rm Drude}(\omega,T) &=& \frac{1}{2}\,{e^2}\alpha\,
\delta(\omega)\left[{\int_{-\infty}^\infty}{dx}\:\:
\frac{\left|\mu+xT \right|\:{e^x}}{\left({e^x}+1\right)^2} \right]\\
\nonumber
\sigma_{xx}^{\rm Inter}(\omega,T) &=& \frac{1}{8}\,{e^2}\alpha\,
\left|{n_F}\left(-\frac{\omega}{2}-\mu\right) -
{n_F}\left(\frac{\omega}{2}-\mu\right)\right|
\end{eqnarray}
where
\begin{equation}
\label{eqn:alpha-def}
\alpha \:=\: \frac{v_F}{v^{\rm DDW}_{\scriptscriptstyle\Delta}}
\:+\: \frac{v^{\rm DDW}_{\scriptscriptstyle\Delta}}{v_F}
\end{equation}
In the zero-temperature limit, this is
\begin{eqnarray}
\sigma_{xx}^{\rm Drude}(\omega,0) &=&
\frac{1}{2}\,{e^2}\alpha\,|\mu|\:\delta(\omega)\cr
\sigma_{xx}^{\rm Inter}(\omega,0) &=& \:\:\frac{1}{8}\,{e^2}\alpha\:\:
\theta\!\left(\left|\frac{\omega}{2}\right|-|\mu|\right)
\end{eqnarray}
The first contribution, $\sigma_{xx}^{\rm Drude}(\omega,T)$,
to the conductivity
is Drude-like: it is proportional to a $\delta$-function in
frequency multiplied by the density of states at
the Fermi level. As in an ordinary Fermi gas,
a uniform electric field can excite a particle-hole
pair of momentum ${\bf q}=0$, which, at $T=0$, must
be precisely at the Fermi surface.
However, the second term, $\sigma_{xx}^{\rm Inter}(\omega,T)$, is special to a
Dirac cone: a uniform, finite-frequency electric field
can excite a quasiparticle from a state of energy
$-\epsilon$ to a state of energy
$\epsilon$ at the same momentum. There is no
gap in the spectrum, so such particle-hole pairs can be created
all the way down to $|\omega|=2|\mu|$.
The existence of this term follows from
both translational symmetry-breaking and rotational
symmetry-breaking. In a crystalline
solid, there will always be, in addition to the
usual Drude term, contributions to
the conductivity resulting from interband transitions.
However, these will be at frequencies higher than the band gap.
In the case of the DDW state, the DDW `band' gap vanishes
at the nodal points as a result of
the $d_{{x^2}-{y^2}}$ order parameter symmetry
which breaks rotational symmetry, so `Interband' transitions contribute
to the conductivity to arbitrarily low frequencies at $\mu=0$.

Note that at $\mu=0$, the Drude contribution
vanishes linearly with temperature,
and the Interband contribution dominates
the conductivity
\begin{eqnarray}
\label{eqn:extreme-cond}
\sigma^{\mu=0}_{xx}(\omega,T) &=& 
\frac{1}{8}\,{e^2}\alpha\,
\left|{n_F}\left(-\frac{\omega}{2}\right) -
{n_F}\left(\frac{\omega}{2}\right)\right| \cr
& & +\,(\ln 2)\,{e^2}\alpha\,T\,
\delta(\omega)
\end{eqnarray}
The Interband contribution extends to zero frequency
and yields a DC conductivity
${e^2}\alpha/8$ at $T=0$.

Let us now consider the thermal conductivity
in the same ideal situation. The thermal
current is not proportional to the momentum even
in a Galilean-invariant system \cite{Nayak01a}, so there
could be a non-Drude contribution -- i.e. one
which is not proportional to $\delta(\omega)$ -- even
in the absence of translational symmetry-breaking.
In the DDW state, this contribution results
from the same excitations which lead to
$\sigma_{xx}^{\rm Inter}(\omega,T)$, so we will use
the same moniker.
\begin{equation}
{\kappa_{xx}}(\omega,T) =  \kappa_{xx}^{\rm Drude}(\omega,T) +
\kappa_{xx}^{\rm Inter}(\omega,T)
\end{equation}
with
\begin{eqnarray}
\kappa_{xx}^{\rm Drude}(\omega,T) &=&\cr
& & {\hskip - 1 cm} \frac{1}{2}\,\alpha \,T\,\delta(\omega)\,
\left[{\int_{-\infty}^\infty}{dx}\, \left|\mu+xT \right|\,
\frac{{x^2}\,{e^x}}{\left({e^x}+1\right)^2} \right]
 \cr
\kappa_{xx}^{\rm Inter}(\omega,T) &=&\cr
& & {\hskip - 1 cm} \frac{1}{4T}\,\alpha\,
{\mu^2} \,\left|{n_F}\left(-\frac{\omega}{2}-\mu\right) -
{n_F}\left(\frac{\omega}{2}-\mu\right)\right|
\end{eqnarray}

At $\mu=0$, the Interband contribution vanishes
and the Drude contribution is proportional
to $T^2$:
\begin{eqnarray}
\label{eqn:extreme-therm-cond}
{\kappa^{\mu=0}_{xx}}(\omega,T) = 
\frac{9}{2}\,\zeta(3)\,\alpha\,{T^2}\,\delta(\omega)\,
\end{eqnarray}
Note that this {\rm strongly} violates the Wiedemann-Franz
law because $\sigma$ and $\kappa/T$ have entirely
different temperature dependences since the former
has an Interband part but the latter does not.

This violation of the Wiedemann-Franz law can be
understood in the following terms. At $\mu=0$,
Interband excitations create a quasihole in a negative
energy state of energy $-{E_k}$ -- therefore costing positive
energy $E_k$ -- and create a
quasiparticle in a positive energy state
with energy ${E_k}=-{E_k}+\omega$. The quasiparticle
and quasihole states have opposite momenta (since the quasihole
is the absence of a particle at ${\bf k}$) and carry
opposite charge. Hence, the resulting state carries
current proportional to $e{\bf k} + (-e)(-{\bf k})=2e{\bf k}$.
However, the two states carry the same energy,
${E_k}=\omega/2$, so the energy current
is ${E_k}{\bf k} + {E_k}(-{\bf k})=0$. Put more
simply, the violation of the Wiedemann-Franz law
is due to the `relativistic' spectrum of the nodal
quasiparticles. If we were to imagine creating
an electron-positron pair with total momentum zero,
we would have the same situation.

For $|\mu|>T$, the Drude contribution is:
\begin{eqnarray}
{\kappa^{\rm Drude}_{xx}}(\omega,T) &=& \frac{\pi^2}{6}\,|\mu|\,
T\,\delta(\omega)
\end{eqnarray}
If $|\mu|<T$, the $\mu=0$ result is recovered.
For $|\omega|\ll|\mu|$, the Interband contribution is
\begin{eqnarray}
\kappa_{xx}^{\rm Inter}(\omega,T) &=&
\frac{{\mu^2}\alpha\,\omega}{16{T^2}\:{\cosh^2}\!\left(\frac{\mu}{2T}\right)}
\sim \frac{1}{T^2}\:{e^{-{\mu}/{2T}}}
\end{eqnarray}
If we try to take $\omega>T$, we
will encounter pathological results, which are related to
heating. In order to obtain sensible answers, we must
include electron-phonon interactions or some other mechanism by
which equilibrium can be maintained. This is analogous
to the divergence which arises in the non-linear
response to an electric field \cite{Tremblay79}.
However, we can consider the limit $\mu\ll\omega\ll T$,
where:
\begin{eqnarray}
\kappa_{xx}^{\rm Inter}(\omega,T) &=&
\frac{{\mu^2}\alpha}{4T}\: \left|\tanh\frac{\omega}{4T} \right|
\end{eqnarray}

Thus, to summarize, we see that, in the limit of a perfectly
clean DDW state, there is no Wiedemann-Franz law
relating $\kappa_{xx}^{\rm Inter}(\omega,T)$ and
$\sigma_{xx}^{\rm Inter}(\omega,T)$. Instead,
\begin{eqnarray}
\kappa_{xx}^{\rm Inter}(\omega,T) = \frac{2 \mu^2}{{e^2}T}\:
\sigma_{xx}^{\rm Inter}(\omega,T)
\end{eqnarray}
Meanwhile, there is a Wiedemann-Franz law
for $\kappa_{xx}^{\rm Drude}(\omega,T)$ and
$\sigma_{xx}^{\rm Drude}(\omega,T)$.
\begin{eqnarray}
\frac{1}{T}\,\kappa_{xx}^{\rm Drude}(\omega,T)=
\frac{L}{e^2}\,\sigma_{xx}^{\rm Drude}(\omega,T)
\end{eqnarray}
where $L={{\pi^2}}/{3}$ for $|\mu|>T$ and
$L={9\,\zeta(3)}/(2\ln 2)$ for $|\mu|<T$.

\subsection{DSC State}

Let us now consider the same quantities in the
DSC state. Because DSC order does not break translational
symmetry, we do not expect to have an Interband
contribution. Furthermore, we might expect the
Drude contribution to vanish at $T=0$ because the chemical potential
merely moves the Dirac points, it does not open Fermi
pockets, unlike in the case of the DDW state. This
is confirmed by a direct calculation:
\begin{equation}
\label{eqn:DSC-cond-clean}
\sigma^{\rm qp}_{xx}(\omega,T) = (\ln 2)\,{e^2}
\left(\frac{v_F}{{v^{\rm DSC}_{\scriptscriptstyle\Delta}}}\right)
T
\end{equation}
At a calculational level, there is a difference
between the DSC case and
the $\mu=0$ DDW case resulting from
the absence of a $\tau^{(3)}$ Pauli matrix in the
current operator in the direction
perpendicular to the bare Fermi surface
in the DSC case. Furthermore, the
current operator in the direction parallel
to the Fermi surface vanishes in the DSC state.
This is a reflection of the coherence factors
of the superconducting state; a quasiparticles
is a superposition of an electron and a hole
which becomes neutral in the limit that the Fermi surface
is approached.

Of course, the superconducting condensate carries
electrical current in the DSC state, so (\ref{eqn:DSC-cond-clean})
does not imply that the state is an electrical
insulator. Note that nodal quasiparticles reduce
the superfluid density in the clean limit, but this
must be obtained from a somewhat different calculation
since the superfluid density is obtained from a different order
of limits of the current-current correlation 
\cite{Scalapino92}.

The thermal conductivity calculation in the DSC
state is identical to that in the DDW state,
with ${v^{\rm DDW}_{\scriptscriptstyle\Delta}}$
replaced by ${v^{\rm DSC}_{\scriptscriptstyle\Delta}}$,
as may be seen from the formal identity between
(\ref{eqn:ddw-energy}) and (\ref{eqn:dsc-energy}).

Thus, we see that, in the clean limit, the quasiparticles
of the DDW and DSC states are very different in their
transport properties despite the similarity in
their spectra. DDW quasiparticles carry electrical
current but negligible thermal current in the
small $\mu$ limit. Even for larger values of
$\mu$, the disparity between electrical and
thermal transport can be seen at finite frequency,
where $\kappa_{xx}^{\rm Inter}(\omega,T)\ll
T\,\sigma_{xx}^{\rm Inter}(\omega,T)$.
On the other hand, the quasiparticles
of the DSC state are able to carry electrical
and thermal currents with essentially equal facility
or lack thereof.

\subsection{Simultaneous DDW and DSC Order}

Finally, we consider a state with simultaneous
DDW and DSC order. The result for the electrical
conductivity is the same as in
a state with DDW order alone, but with a modified
$\alpha$:
\begin{equation}
\label{eqn:mod-alpha}
\alpha \:\rightarrow\: \frac{1}{\sqrt{\left(
{v^{\rm DDW}_{\scriptscriptstyle\Delta}}\right)^2
+ \left({v^{\rm DSC}_{\scriptscriptstyle\Delta}}\right)^2}}\,
\left({v_F}
\:+\: \frac{\left(
{v^{\rm DDW}_{\scriptscriptstyle\Delta}}\right)^2}{v_F}\right)
\end{equation}
The conductivity has both Drude and Interband
components because translational symmetry is broken by the
DDW order parameter. The density-of-states is decreased
by the development of DSC order, and the conductivity
is decreased by this factor.

The thermal conductivity is the same as in the
DDW or DSC state, but with ${v^{\rm DDW}_{\scriptscriptstyle\Delta}}$
or ${v^{\rm DSC}_{\scriptscriptstyle\Delta}}$
replaced by:
\begin{equation}
\label{eqn:mod-v}
{v^{\rm DDW}_{\scriptscriptstyle\Delta}},\:\:
{v^{\rm DSC}_{\scriptscriptstyle\Delta}} \rightarrow
{\sqrt{\left(
{v^{\rm DDW}_{\scriptscriptstyle\Delta}}\right)^2
+ \left({v^{\rm DSC}_{\scriptscriptstyle\Delta}}\right)^2}}
\end{equation}

\section{Electrical and Thermal Conductivities in the
Impurity-Scattering-Dominated Limit}

Now consider the more realistic situation in
which the quasiparticles of the DDW state have finite
lifetimes as a result of impurity scattering.
We will continue to neglect the effects of
quasiparticle-quasiparticle scattering, which
is irrelevant in the low-frequency, low-temperature limit.
(The primary effect of electron-electron interactions
is, thus, assumed to be the formation of
the DDW state; the residual interactions
can be neglected in the low-energy limit.)
The basic effect of impurity scattering is to smear
out momenta so that $E(k)$ has an uncertainty
$1/\tau$. As a result, the frequency dependence
of our expressions for the electrical and thermal
conductivity are essentially convoluted with
a Lorentzian in frequency of width $1/\tau$.
Thus, for instance, $\delta(\omega)$ is replaced by
$(\tau/\pi)/(1+{\omega^2}{\tau^2})$. Furthermore, we
expect that $\mu$ will be replaced by a function
which behaves like $\text{max}(\mu,1/2\tau)$. Note that
when $\tau$ is finite, there is no longer a sharp
separation between the Drude and Interband components
of the conductivity. Nevertheless,
we will still use this terminology
since, as a practical matter, we can decompose
the conductivity into the impurity-broadened versions
of the two components of the clean limit.

Consider the DC conductivity of the
DDW state for $\mu=0$:
\begin{eqnarray}
\nonumber
\sigma^{DC}_{xx}(0,T) &=& 2\pi
\int \frac{{d^2}k}{\left(2\pi\right)^2}\:\text{Tr}\!\left\{
{A^B}(k,0)\,\tau^{(3)}\,{A^B}(k,0)\,\tau^{(3)}\right\}\\
&=& \frac{e^2}{2\pi^2}\,\alpha
\end{eqnarray}
From the expression for $A(k,0)$ given in the appendix,
we see that it is proportional to the identity
matrix. Hence, the $\tau^{(3)}$s have no effect. As a result,
the DC conductivity of quasiparticles in the DSC
state follows from an identical calculation,
with ${v^{\rm DDW}_{\scriptscriptstyle\Delta}}$
replaced by ${v^{\rm DSC}_{\scriptscriptstyle\Delta}}$
but there is only a contribution
from the single pair of nodes at which
${j_x}\neq 0$ in the DDW state:
\begin{eqnarray}
\sigma^{DC}_{xx}(0,T) &=&  \frac{e^2}{2\pi^2}\,
\frac{v_F}{v^{\rm DSC}_{\scriptscriptstyle\Delta}}
\end{eqnarray}
As we discussed above, the Drude contribution
to the conductivity is modified by two effects:
the Dirac points are broadened and aquire a width
$1/\tau$; and the $\delta$-function is broadened
into a Lorentzian with zero-frequency height
$\tau$. The $\tau$-dependence cancels, so
the Drude term in the
conductivity is independent of $\tau$.

Now consider the DDW state for finite $\mu$.
If $|\mu|>1/\tau$, the conductivity in
the zero-temperature DC limit is
\begin{eqnarray}
\sigma_{xx}^{\rm Drude}(0,0) &=&
\frac{1}{4\pi}\,{e^2}\alpha\,|\mu|\tau
\end{eqnarray}
There is only a Drude component in this
limit.

\begin{figure*}[th]
\epsfysize=5cm 
\centerline{\epsfbox{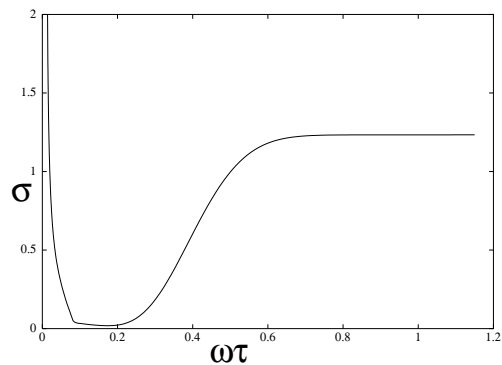}}
\vskip 0.5 cm
\caption{The frequency-dependent conductivity
of the DDW state in the impurity-scattering-dominated
limit. The conductivity is in units of $8/{e^2}\alpha$
while the scale for $\omega\tau$ is $10^{-4}$.}
\label{fig:cond-w-tau}
\end{figure*}

Now let us consider the Interband
conductivity for finite $\tau$. The DSC
state has no such contribution, whether
or not $\tau$ is finite. Hence, we consider
the DDW state only. The Interband contribution
is qualitatively the same as in the clean limit,
but now smeared on a frequency scale $1/\tau$,
as may be seen from the plot of conductivity
vs. frequency in Fig. \ref{fig:cond-w-tau}

The thermal conductivity is formally
identical in the DSC state and the DDW state at
$\mu=0$. The Drude contribution is:
\begin{eqnarray}
\kappa_{xx}^{\rm Drude}(0,T\rightarrow 0) &=&
\frac{1}{6}\,\alpha \,T\,
\end{eqnarray}
where $\alpha$ is understood to be the
expression in (\ref{eqn:alpha-def}) or its analogue
for the DSC state.

\begin{figure*}[th]
\epsfysize=5cm 
\centerline{\epsfbox{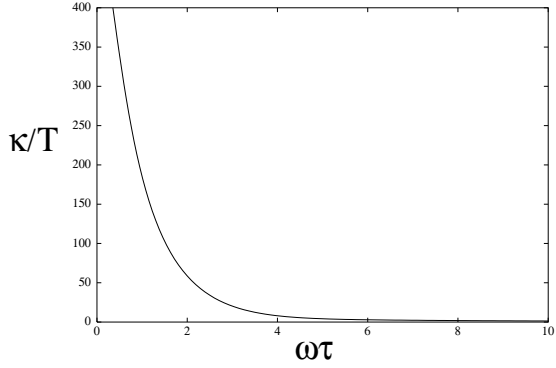}}
\vskip 0.5 cm
\caption{The frequency-dependent thermal conductivity
of the DDW state in the impurity-scattering-dominated
limit. The thermal conductivity divided by $T$
is in units of $8/{e^2}\alpha$
while the scale for $\omega\tau$ is $10^{-1}$.}
\label{fig:tcond-w-tau}
\end{figure*}

The Interband contribution, which vanishes
in the clean limit for $\mu=0$ is now non-vanishing,
though small, as a result of the impurity-induced
smearing on a frequency scale $1/\tau$,
as may be seen from the plot of thermal conductivity
vs. frequency in Fig. \ref{fig:tcond-w-tau}

The contrast between Figs. \ref{fig:cond-w-tau}
and \ref{fig:tcond-w-tau}
summarizes the distinction between the
electrical and thermal conductivities of
DDW nodal quasiparticles. In the case of the DSC state,
there is no such contrast.

The basic fact which we uncovered in the $\tau=\infty$
limit still holds: nodal quasiparticles in the DDW state
are qualitatively better carriers of electrical current
than thermal current while nodal quasiparticles
in the DSC state exhibit no such dichotomy.

As in the clean case, the electrical conductivity in
a state with both DDW and DSC order follows the
DDW case, with both a Drude and an Interband part.
The main difference is the modified $\alpha$ of
(\ref{eqn:mod-alpha}). The thermal conductivity
is the same as in either the DDW or DSC states,
but with the velocities combined as in (\ref{eqn:mod-v}).

\section{Electrical and Thermal Hall Conductivities}

We now consider transport in a magnetic field.
For weak magnetic fields, we would, in general,
expect $\sigma_{xy}, \kappa_{xy}\propto B$
purely on symmetry grounds. A direct calculation
shows that in the DDW state,
\begin{eqnarray}
\sigma_{xy} &=& 16e^{3}B\tau^{2}v_{F}v^{\rm DDW}_{\scriptscriptstyle\Delta} \\
\kappa_{xy} &=& \frac{16\pi^{2}}{3} eB\tau^{2}
Tv_{F}v^{\rm DDW}_{\scriptscriptstyle\Delta}
\end{eqnarray}
Note that this result holds for finite $\mu$:
the $\mu$ dependence of the hole density, $|\mu|$,
cancels the $\mu$ dependence of the cyclotron frequency,
$ev_{F}v^{\rm DDW}_{\scriptscriptstyle\Delta}/|\mu|$.
Hence, the Hall constant is
\begin{eqnarray}
{R_H} =  \frac{256}{\pi^2}\:
\frac{{v_F}\,v^{\rm DDW}_{\scriptscriptstyle\Delta}}{\alpha^2}
\,\, \text{min}\!\left(\frac{1}{|\mu|^2},{\tau^2}\right)
\end{eqnarray}
The effective carrier concentration measured by
$R_H$ varies as $\mu^2$ -- not as $\mu$ -- as a result
of the peculiar $\mu$-independence of $\sigma_{xy}$.
Keep in mind, however, that these results are strictly for
the regime in which transport is dominated by nodal
quasiparticles.

The large $B$ limit can be understood by considering
the effect of a magnetic field on nodal quasiparticles in
the clean limit. Nodal quasiparticles satisfy the equation
\begin{eqnarray}
\left[{\sigma^{(3)}_{AB}} {v_F} \left(i {\partial_x}+e{A_x}\right) 
\:+\:{\sigma^{(1)}_{AB}} {v_{\scriptscriptstyle\Delta}}
\left(i {\partial_y}+e{A_y}\right)\right] {\chi_{\alpha A}}
=\cr E\,{\chi_{\alpha A}}
\end{eqnarray}
Squaring the differential operator which acts on $\chi$,
we have:
\begin{eqnarray}
\biggl[{v_F^2} {\left(i {\partial_x}+e{A_x}\right) ^2}
+ {v_{\scriptscriptstyle\Delta}^2}
{\left(i {\partial_y}+e{A_y}\right)^2} {\hskip 2 cm}\cr
 +\,
\sigma^{(2)}{v_F}{v_{\scriptscriptstyle\Delta}}eB\biggr] {\chi}
=  {E^2}\,{\chi}
\end{eqnarray}

The solutions of this equation form Landau levels at energies
\begin{equation}
{E_n} = \pm
\sqrt{{v_F}{v_{\scriptscriptstyle\Delta}}eB\:n\:  } 
\end{equation}
for $n=0,1,\ldots$.
For ${\omega_c}\tau \gg 1$, these will give quantized
Hall conductances.

By way of contrast, consider the linear response calculation
in the DSC state and in the state with DDW and DSC order
\begin{eqnarray}
\Pi_3^{CCC}(i{\omega_n}\rightarrow\omega+i\delta,{\bf q},T) &=& 0\cr
\Pi_3^{EEC}(i{\omega_n}\rightarrow\omega+i\delta,{\bf q},T) &=& 0
\end{eqnarray}
Consequently, the electrical and thermal Hall
conductivities vanish in these states {\em at
the level of approximation used in this paper},
namely linearization about the Fermi surface. In fact,
they vanish {\em a fortiori} in this approximation,
since the quasiparticles in
a superconducting state do not couple to the electromagnetic
field ${\bf A}$ alone, but to the supercurrent
${\bf\nabla}\varphi-{\bf A}$, so that these response functions
must be multiplied by the curl of the supercurrent -- which
oscillates about zero mean in the vortex lattice state --
not the magnetic field.

A number of authors have shown that non-zero
Hall conductivities are obtained in the superconducting
state only when one takes into account corrections
to the linearization about the nodes. This
is discussed thoroughly elsewhere, so we
will not dwell on it here, and will limit
ourselves to the observation that the situation
in the DDW state is very different from that
in the DSC state. According to the scaling laws
discussed in \cite{Simon97,Ye01,Vafek01},
in the DSC state
\begin{equation}
\kappa_{xy} \sim T \sqrt{B}
\end{equation}
Furthermore, the nodal quasiparticles of
the DSC state do not form Landau levels but,
rather, Bloch waves in the periodic potential
caused by the supercurrent \cite{Franz00}.

\section{Discussion}

Nodal quasiparticles are created equal, but some are more equal
than others. The nodal quasiparticles
of the DDW state are clearly different from those of the DSC state.
The differences are a reflection of the symmetries which are broken
in the two states. Broken translational symmetry in the DDW state
leads to the existence of an Interband component of the electrical
conductivity. This is a manifestation of the fact
that quasiparticles and quasiholes in the
DDW state can be visualized as electrons and positrons.
A quasiparticle and quasihole with equal and opposite momenta
carry twice the electrical current carried by either one of them
and, at $\mu=0$, zero energy current. These properties of
DDW quasiparticles follow from the
translational and rotational symmetry of the state, so
they also apply to quasiparticles in graphite, in some models
of quantum critical points \cite{Ludwig94},
and in multi-band models \cite{Varma99}.

This divergence between electrical
and thermal transport is reminiscent of that observed
in the experiments of \cite{Hill01}, where
${\kappa}/\sigma T$ is found vanish at low temperatures
in Pr$_{2-x}$Ce$_x$CuO$_4$ at magnetic fields greater then
the upper critical field -- i.e. in the presumed low-temperature
`normal state' with superconductivity suppressed. Such
a severe violation only occurs in the DDW state
in the extreme limit $\mu=0$, $\tau=\infty$, as may
be seen from (\ref{eqn:extreme-cond}) and
(\ref{eqn:extreme-therm-cond}). Nevertheless,
it is interesting that such a counter-intuitive
violation is possible at all.

There is no Interband component in the DSC state since it
does not break translational symmetry.
The DSC state has vanishing electrical and thermal Hall
conductivities at the linear response level because
$U(1)$ gauge symmetry is broken to $Z_2$ by the condensation
of a charge-$2e$ order parameter. A DSC quasiparticle
is a mixture of an electron and a hole, which should not have
a Hall effect.  Quasiparticles of the DSC state are
essentially superpositions of particle
and hole excitations of an ordinary Fermi surface,
and they inherit the familiar properties of such excitations:
they carry ${\pi^2}T/3$ units of energy for each
unit $e$ of charge. (However, in any superconductor,
the condensate carries electrical current but no thermal current, again
violating the Wiedemann-Franz law.)

At present, there is insufficient experimental
data on clean, highly-underdoped cuprates with a well-developed
pseudogap to make a comparison with our results.
This is the relevant limit because
our calculations are applicable to the regime
in which transport is dominated by quasiparticles which
are close enough to the nodal points $(\pm\frac{\pi}{2a}, \pm\frac{\pi}{2a})$
that we can safely consider only linear order
in deviations of the momentum about these points.
In other words, our results apply to small values
of the chemical potential $\mu$.
They would certainly apply to a putative DDW state
very close to half-filling,
but DDW order is unlikely to occur very close to
half-filling since antiferromagnetism is the
predominant ordering tendency there. At the
more significant doping levels at which superconductivity
occurs, we might worry that the approximations used in this
paper will no longer apply to the DDW state. However,
there are several reasons why we believe that these
fears may be misplaced. Angle-resolved photoemission
experiments find a Fermi surface which has not
moved very far along the Brillouin zone diagonals
from $(\pm\frac{\pi}{2a}, \pm\frac{\pi}{2a})$ \cite{Norman98,Harris96}.
Even a naive determination
of the Fermi surface, using the band structure of
(\ref{eqn:band-structure}) and (\ref{eqn:gap-structure})
with $t'=0.45\,t$ finds small hole pockets
about $(\pm\frac{\pi}{2a}, \pm\frac{\pi}{2a})$.
Furthermore, measurements of the chemical potential
\cite{Ino97} find that $\mu\propto x^2$ rather
than the naively-expected dependence $\mu\propto x$;
a similar result is found in Monte Carlo numerical studies of
the $2D$ Hubbard model \cite{Dagotto91,Furukawa93}. 
This could occur if the doped holes are confined to the boundaries
between hole-poor regions \cite{Zaanen96,Emery97},
thereby pinning the chemical potential
close to zero. Thus, our finite-frequency results,
which are most interesting in the regime $\mu<\omega<{\Delta_0}$
may have a non-zero window of validity. We find it curious
and possibly quite fortuitous
that high quality YBCO crystals, which find
evidence for DDW order, also exhibit the strongest evidence
for dynamical stripes \cite{Mook99}, which could pin the chemical
potential at small values.

To summarize, we have, in this paper, shown how the
differences between nodal quasiparticles in
the DDW and DSC states can
be probed by low-temperature transport measurements.
The most notable signatures are the dichotomy between
electrical and thermal transport
and the existence of a linear in $B$ Hall response
in the DDW state -- neither of which is a characteristic
of nodal quasiparticles in the DSC state.

We would like to thank Sudip Chakravarty for
discussions.
This work was supported by the National Science Foundation under
Grant No. DMR-9983544 and the A.P. Sloan Foundation.

\appendix

\section{Kubo Formulae}

The Kubo formulae relate transport coefficients
to retarded correlation functions of current operators.
We will obtain the latter by analytic continuation
of the imaginary-time correlation functions.

To this end, we define the Matsubara-frequency
current-current correlation functions:
\begin{eqnarray}
\Pi_2^{CC}(i{\omega_n},{\bf q},T) &=&
{\int_0^\beta} d\tau \,{e^{i{\omega_n} \tau}}\, \left\langle {T_\tau}
{j_x}({\bf q},\tau){j_x}(-{\bf q},0)
\right\rangle\cr
\Pi_2^{EE}(i{\omega_n},{\bf q},T) &=&
{\int_0^\beta} d\tau \,{e^{i{\omega_n} \tau}}\, \left\langle {T_\tau}
{j^E_x}({\bf q},\tau){j^E_x}(-{\bf q},0)
\right\rangle\cr
\Pi_2^{EC}(i{\omega_n},{\bf q},T) &=&
{\int_0^\beta} d\tau \,{e^{i{\omega_n} \tau}}\, \left\langle {T_\tau}
{j^E_x}({\bf q},\tau){j_x}(-{\bf q},0)
\right\rangle
\end{eqnarray}
and three-current correlation functions:
\begin{eqnarray}
\Pi_3^{CCC}(i{\omega_n},{\bf q},T) &=&\cr
& & {\hskip -2.5 cm}{\int_0^\beta} d\tau\, d\tau'\,\,{e^{i{\omega_n} \tau}}\,
\left\langle {T_\tau}{j_y}({\bf q},\tau)
{j_x}(0,0){j_y}(-{\bf q},\tau')
\right\rangle\cr
\Pi_3^{ECC}(i{\omega_n},{\bf q},T) &=&\cr
& & {\hskip -2.5 cm}
{\int_0^\beta} d\tau\, d\tau'\,\,{e^{i{\omega_n} \tau}}\,
\left\langle {T_\tau}{j^E_y}({\bf q},\tau)
{j_x}(0,0){j_y}(-{\bf q},\tau')
\right\rangle\cr
\Pi_3^{EEC}(i{\omega_n},{\bf q},T) &=&\cr
& & {\hskip -2.5 cm}
{\int_0^\beta} d\tau\, d\tau'\,\,{e^{i{\omega_n} \tau}}\,
\left\langle {T_\tau}{j^E_y}({\bf q},\tau)
{j^E_x}(0,0){j_y}(-{\bf q},\tau')
\right\rangle
\end{eqnarray}

The frequency-dependent electrical conductivity is given
by the Kubo formula
\begin{eqnarray}
\sigma_{xx}(\omega,T)
= \frac{1}{\omega}\,\text{Im}\left\{ 
\Pi_2^{CC}(i{\omega_n}\rightarrow\omega+i\delta,0,T)\right\}
\end{eqnarray}
Meanwhile, the Hall conductivity is obtained from
a correlation function of three currents,
\begin{eqnarray}
\sigma_{xy}(\omega,T)/B &=&\cr
& & {\hskip -1 cm}
 \frac{1}{\omega}\,\frac{1}{q_x}\,\text{Im}\left\{
\Pi_3^{CCC}(i{\omega_n}\rightarrow\omega+i\delta,{\bf q}\rightarrow 0,T)
\right\}
\end{eqnarray}

The thermal conductivity is given by the
the following combination of correlation functions
\begin{eqnarray}
\kappa_{xx}(\omega,T) &=& 
\frac{1}{T}\,\frac{1}{\omega}\,
\text{Im}\left\{
\Pi_2^{EE}(i{\omega_n}\rightarrow\omega+i\delta,0,T)\right\}\cr
& &-\: T\,{S^2}(\omega,T) \sigma_{xx}(\omega,T)
\end{eqnarray}
The second term on the right-hand-side ensures
that the energy current flows under a condition
of vanishing electrical current. It involves the
thermopower, $S(\omega,T)$:
\begin{eqnarray}
S(\omega,T)  &=& 
-\,\frac{1}{T}\,\frac{\text{Im}\left\{
\Pi_2^{EC}(i{\omega_n}\rightarrow\omega+i\delta,0,T)\right\}}
{\text{Im}\left\{
\Pi_2^{CC}(i{\omega_n}\rightarrow\omega+i\delta,0,T)\right\}}
\end{eqnarray}
The thermal Hall conductivity is given by
\begin{eqnarray}
\kappa_{xy}(\omega,T)/B &=&\cr
& & {\hskip -1.5 cm}
 \frac{1}{T}\,\frac{1}{\omega}\,\frac{1}{q_x}\,\text{Im}\left\{
\Pi_3^{EEC}(i{\omega_n}\rightarrow\omega+i\delta,{\bf q}\rightarrow 0,T)
\right\}\cr
& &-\: T\,{N^2}(\omega,T) \sigma_{xy}(\omega,T)
\end{eqnarray}
where $N(\omega,T)$ is the Nernst coefficient,
or Hall thermopower:
\begin{eqnarray}
N(\omega,T)/B &=&\cr
& & {\hskip -1.5 cm}
-\,\frac{1}{T}\,\frac{\text{Im}\left\{
\Pi_2^{ECC}(i{\omega_n}\rightarrow\omega+i\delta,
{\bf q}\rightarrow 0,T)\right\}}
{\text{Im}\left\{
\Pi_2^{CCC}(i{\omega_n}\rightarrow\omega+i\delta,
{\bf q}\rightarrow 0,T)\right\}}
\end{eqnarray}

\section{Impurity Scattering of Nodal Quasiparticles}

We briefly review the effect of impurity scattering
on nodal quasiparticles. Let us assume that the impurities
can be described by a $\delta$-function-correlated,
Gaussian-distributed random potential with
variance $V_0$:
\begin{equation}
\overline{V({\bf x})\,V({\bf x'})} = {|V_0|^2}\,
\delta^{(2)}({\bf x}-{\bf x'})
\end{equation}

\begin{figure*}[th]
\epsfysize=1cm 
\centerline{\epsfbox{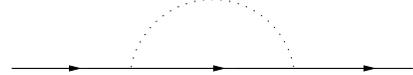}}
\vskip 0.5 cm
\caption{The leading diagram which contributes to the quasiparticle
self-energy.}
\label{fig:self-energy}
\end{figure*}

Consider the graph in figure (\ref{fig:self-energy})
which determines the quasiparticle self-energy
to lowest order. For nodal quasiparticles,
both the real and imaginary parts
are logarithmically divergent in the infrared.
Hence, we compute the self-energy self-consistently
\begin{equation}
{\Sigma^1}\left(i{\epsilon_n}\right) =
\frac{|V_0|^2}{{v_F}v_{\scriptscriptstyle\Delta}}
\int \frac{{d^2}k}{(2\pi)^2}\,
\frac{i{\epsilon_n}+\Sigma\left(i{\epsilon_n}\right)+
{\bf\sigma}\cdot{\bf k}}{{\left({\epsilon_n}+
i{\Sigma^1}\left(i{\epsilon_n}\right)\right)^2} + k^2}
\end{equation}
A similar expression holds for the self-energy for
the other pair of nodes, ${\Sigma^2}\left(i{\epsilon_n}\right)$.
In this expression, we have rescaled ${k_x},{k_y}$
to remove the velocities from the Green function
and bring them outside the integral.
Solving this equation self-consistently, we
find ${\Sigma^1}\left(i{\epsilon_n}\rightarrow 0\right)
= i/2\tau\:\text{sgn}({\epsilon_n})$)
with
\begin{equation}
\tau = \frac{1}{\Lambda}\,
{e^{2\pi{{v_F}v_{\scriptscriptstyle\Delta}}/{|V_0|^2}}}
\end{equation}
where $\Lambda$ is the ultraviolet cutoff on the
nodal effective Lagrangian. This effective
theory certainly breaks down at scales comparable
to the maximum of the gap, so we should take
$\Lambda\stackrel{<}{\scriptscriptstyle\sim} \Delta_0$.
Meanwhile,
\begin{equation}
\text{Re}{\Sigma^1}\left(i{\epsilon_n}\rightarrow 0\right) = \left[
\frac{\frac{|V_0|^2}{2\pi{{v_F}v_{\scriptscriptstyle\Delta}}}
\ln(2\Lambda\tau)}{1 -
\frac{|V_0|^2}{2\pi{{v_F}v_{\scriptscriptstyle\Delta}}}\ln(2\Lambda\tau)}
\right]\: i{\epsilon_n}
\end{equation}
This can be absorbed into a redefinition of the velocities.

Hence, the Green function in the presence of
impurities is:
\begin{eqnarray}
\label{eqn:imp-Green}
{G^1}({i\epsilon_n},{\bf k}) = \frac{i{\epsilon_n}+
i/2\tau\:\text{sgn}({\epsilon_n})
+ {\bf \sigma}\cdot{\bf k}}{{({i\epsilon_n}
+i/2\tau\:\text{sgn}({\epsilon_n}))^2}-{k^2}}
\end{eqnarray}
where the momenta have been rescaled so that
${k^2}={v_F^2}{k_x^2}~+~v^2_{\scriptscriptstyle\Delta}{k_y^2}$
and ${\bf k}=\left({v_F}{k_x},v_{\scriptscriptstyle\Delta}{k_y}\right)$
on the right-hand-side of (\ref{eqn:imp-Green}).

This result holds for the DDW state at $\mu=0$
or the DSC state for arbitrary $\mu$. In the former case, the
matrix indices correspond to momenta ${\bf k}$ and ${\bf k+Q}$;
in the latter case, they are particle/hole indices.
The associated matrix spectral function $A(k,0)$ is
\begin{eqnarray}
{A^1}(\epsilon,{\bf k}) &=&\cr
& &  {\hskip - 1 cm} \frac{1}{2\pi\tau}\,
\frac{{\epsilon^2}+{k^2} + {1/(2\tau)^2} +
2\epsilon{\bf \sigma}\cdot{\bf k}}{
\left[{(\epsilon+k)^2} + {1/(2\tau)^2}\right]
\left[{(\epsilon-k)^2} + {1/(2\tau)^2}\right]
}
\end{eqnarray}
For the DDW state at non-zero $\mu$, we have
\begin{eqnarray}
{A^1}(\epsilon,{\bf k}) &=&\cr
& & {\hskip - 1.5 cm} \frac{1}{2\pi\tau}\,
\frac{{(\epsilon+\mu)^2}+{k^2} + {1/(2\tau)^2} +
2(\epsilon+\mu){\bf \sigma}\cdot{\bf k}}{
\left[{(\epsilon+\mu+k)^2} + {1/(2\tau)^2}\right]
\left[{(\epsilon+\mu-k)^2} + {1/(2\tau)^2}\right]
}
\end{eqnarray}
with similar expressions for ${A^2}(\epsilon,{\bf k})$.

These spectral functions are used in the calculations
in sections VI and VII.


\begin{references}

\bibitem{Ginsberg89} For reviews, see {\it Physical Properties
of High Temperature Superconductors}, Edited by D. M. Ginsberg
 (World Scientific, Singapore) Vol. I (1989), Vol. II (1990),
Vol. II (1992), Vol. IV (1994).

\bibitem{Chakravarty01} S. Chakravarty, R. B. Laughlin, D. K. Morr,
C. Nayak, Phys. Rev. B {\bf 63}, 94503 (2001).

\bibitem{Timusk99} T. Timusk and B. Statt,
Rep. Prog. Phys. {\bf 62}, 61 (1999).

\bibitem{Nayak00a} C. Nayak, Phys. Rev. B {\bf 62}, 4880 (2000).

\bibitem{Norman98} M. R. Norman et al., Nature {\bf 392}, 157(1998).

\bibitem{Harris96} J. M. Harris, {\it et al.}, Phys. Rev.
B {\bf 54}, R15665 (1996).

\bibitem{Renner98} Ch. Renner {\it et al.}, Phys. Rev. Lett. {\bf 80},
149(1998).

\bibitem{Fong00} H. F. Fong, {\it et al.},  Phys. Rev.
B {\bf 61}, 14773 (2000).

\bibitem{Dai99} P. Dai, {\it et al.}, Science {\bf 284},
1344 (1999).

\bibitem{Homes93} C. C. Homes, {\it et al.}, Phys. Rev. Lett.
{\bf 71}, 1645 (1993).

\bibitem{Tallon99} J. L.Tallon {\it et al.}, Phys. Stat. Sol. B {\bf 215},
531 (1999); J. W. Loram {\it et al.}, J. Phys. Chem. Solids
{\bf 59}, 2091 (1998).

\bibitem{Bilbro76} G. Bilbro and W. L. McMillan,
Phys. Rev. B {\bf 14}, 1887 (1976).

\bibitem{Vescoli98} V. Vescoli, {\it et al.}, Phys. Rev. Lett.
{\bf 81}, 453 (1998).

\bibitem{Neto00} A. H. Castro Neto, cond-mat/0012147.

\bibitem{Mook01} H. A. Mook, P. Dai, F. Dogan, Phys. Rev. B
{\bf 64}, 012502 (2001); H. A. Mook, {\it et al.}, unpublished.

\bibitem{Sonier} J. E. Sonier, {\it et al.},
Science {\bf 292}, 1692 (2001).

\bibitem{Nayak01a} C. Nayak, {\it et al.}, cond-mat/0105357.

\bibitem{Scalapino92} D. J. Scalapino, S. R. White,
and S. C. Zhang, Phys. Rev. Lett. {\bf 68}, 2830 (1992).

\bibitem{Tremblay79} A.-M. Tremblay, {\it et al.},
Phys. Rev. A {\bf 19}, 1721 (1979).

\bibitem{Simon97} S. H. Simon and P.A. Lee,
Phys. Rev. Lett. {\bf 78}, 1548 (1997).

\bibitem{Ye01} J. Ye, Phys. Rev. Lett. {\bf 86}, 316 (2001).

\bibitem{Vafek01} O. Vafek, {\it et al.}, cond-mat/0104516.

\bibitem{Franz00} M. Franz and Z. Tesanovic, Phys. Rev.
Lett, {\bf 83}, 554 (2000).

\bibitem{Ludwig94}
A. Ludwig, M. P. A.  Fisher, R. Shankar, and
G. Grinstein, Phys. Rev. B {\bf 50}, 7526 (1994).

\bibitem{Varma99}
C.~M. Varma, Phys. Rev. Lett. {\bf 83}, 3538 (1999).

\bibitem{Ino97} A. Ino {\it et al.}, Phys. Rev. Lett, {\bf 79}, 2101 (1997).

\bibitem{Dagotto91} E. Dagotto {\it et al.}, Phys. Rev. Lett, {\bf 67},
1918 (1991).

\bibitem{Furukawa93} N. Furukawa and M. Imada, J. Phys. Soc. Jpn. {\bf 62},
2557 (1993).

\bibitem{Zaanen96} J. Zaanen and A. M. Ol\'e, Ann. Phys. (Leipzig)
{\bf 5}, 224 (1996).

\bibitem{Emery97} V. J. Emery, S. A. Kivelson, and O. Zachar, Phys. Rev. B, 
{\ bf 56}, 6120, (1997).

\bibitem{Mook99} H. A. Mook and F. Dogan, Nature {\bf 401},
145 (1999); P. Dai, H. A. Mook, R. D. Hunt, and F. Dogan,
Phys. Rev. B {\ bf 63}, 54525 (2001).

\bibitem{Hill01} R.W. Hill, {\it et al.}, unpublished.


\end{references}
\end{document}